\documentclass{emulateapj}
\usepackage{natbib}
\usepackage{bm}
\usepackage{mathrsfs,amsmath}
\usepackage{graphicx}
\usepackage{epstopdf}
\usepackage[english]{babel}
\usepackage{xcolor}
\usepackage[colorlinks,linkcolor={blue},citecolor={blue},urlcolor={blue}]{hyperref}

\newcommand{\yam}[1]{\textcolor{blue}{Yam13}}
\defcitealias{yam13}{Yam13}

\shorttitle{The progenitor of the Cetus stream}
\shortauthors{Yuan et al.}

\begin{document}

\title{Revealing the complicated story of the Cetus Stream with StarGO}


\author{Zhen Yuan\altaffilmark{1}}
\author{M. C. Smith\altaffilmark{1}}

\author{Xiang-Xiang Xue\altaffilmark{2}}
\author{Jing Li\altaffilmark{3,1}}

\author{Chao Liu\altaffilmark{2}}
\author{Yue Wang\altaffilmark{2}}
\author{Lu Li\altaffilmark{1,4}}
\author{Jiang Chang\altaffilmark{2,5}}

\altaffiltext{1}{Key Laboratory for Research in Galaxies and Cosmology, Shanghai Astronomical Observatory, Chinese Academy of Sciences, 80 Nandan Road, Shanghai 200030, China; sala.yuan@gmail.com, lijing@shao.ac.cn}
\altaffiltext{2}{Key Laboratory of Optical Astronomy, National Astronomical Observatories, Chinese Academy of Sciences, Beijing 100101, China}
\altaffiltext{3}{Physics and Space Science College, China  West Normal University, 1 ShiDa Road, Nanchong 637002, China}

\altaffiltext{4}{University of the Chinese Academy of Sciences, No.19A Yuquan Road, Beijing 100049, China}

\altaffiltext{5}{Purple Mountain Observatory, CAS, No.8 Yuanhua Road, Qixia District, Nanjing 210034, China}

\begin{abstract}

We use a novel cluster identification tool \textsc{StarGO} to explore the metal poor ([Fe/H] $<$ -1.5) outer stellar halo (d $>$ 15 kpc) of the Milky Way using data from Gaia, LAMOST and SDSS. Our method is built using an unsupervised learning algorithm, a self-organizing map, which trains a 2-D neural network to learn the topological structures of a data set from an n-D input space. Using a 4-D space of angular momentum and orbital energy, we identify three distinct groups corresponding to the Sagittarius, Orphan, and Cetus Streams. For the first time we are able to discover a northern counterpart to the Cetus stream. We test the robustness of this new detection using mock data and find that the significance is more than 5-sigma. We also find that the existing southern counterpart bifurcates into two clumps with different radial velocities. By exploiting the visualization power of \textsc{StarGO}, we attach MW globular clusters to the same trained neural network. The Sagittarius stream is found to have five related clusters, confirming recent literature studies, and the Cetus stream has one associated cluster, NGC 5824. This latter association has previously been postulated, but can only now be truly confirmed thanks to the high-precision Gaia proper motions and large numbers of stellar spectra from LAMOST. The large metallicity dispersion of the stream indicates that the progenitor cannot be a globular cluster. Given the mean metallicity of the stream, we propose that the stream is the result of a merger of a low-mass dwarf galaxy that is associated to a massive globular cluster (NGC 5824).

\end{abstract}

\keywords{galaxies: halo --- galaxies: kinematics and dynamics --- galaxies: formation --- methods: data analysis}

\section{Introduction}
\label{intro}


The Milky Way is expected to go through frequent merger events with nearby dwarf satellite galaxies during its assembly history. Each merger event will leave streams in the stellar halo, the kinematic information of which encodes the disruption history of their progenitor systems. According to the hierarchical structure formation paradigm, in general streams that stay in the outer stellar halo come from systems accreted later compared to those in the inner region \citep{bullock05, cooper10, amorisco17}. Moreover, the effects of both tidal disruption and dynamical friction are milder as the Galacto-centric distance increases. Thus the streams in the outer stellar halo are less phase mixed and the clustering features should be better preserved in phase space. However, the identification of these distant streams has previously been hindered by the lack of suitable proper motions. This situation is now changing thanks to the Gaia survey \citep{gaia16,gaia18}, which is delivering precise proper motions across the entire sky down to $\sim$21st magnitude. Accompanied by large spectroscopic surveys such as SDSS \citep{york00} and LAMOST \citep{cui12, zhao12,deng12}, we are now able to search for halo substructures using precise 6-D phase space for the first time. 

The most significant halo substructure found in the modern era is the disrupted Sagittarius dwarf galaxy (\citealt{ibata94, ibata95, yanny00}; hereafter Sgr), and its associated streams \citep{mateo96, ibata01, maj03}. Near the Sgr trailing arm there is another stream, named the Cetus Polar Stream (CS), which was first discovered by \citet{newberg09} using photometry and spectra from SDSS/SEGUE data release 7 \citep{sdss7}. They also found that a distant halo globular cluster (GC), NGC 5824, is likely to be associated with CS, using an orbit deduced from stream's radial velocity. \citet{yam13}, hereafter \yam, updated the results of the distance and velocity measurement of CS using SDSS data release 8, including N-body simulation to show that CS can be reproduced by a disrupted dwarf galaxy with mass of $10^{8}$M$_{\odot}$.

Prompted by the association between NGC 5824 and the CS, a number of studies have investigated the nature of this GC to test whether it may be the disrupted core of the progenitor. \citet{costa14} find there is an intrinsic metallicity dispersion around 0.3 dex from low resolution spectroscopy of Ca II triplet absorption lines of 118 red giant members. However, a separate study using a restricted sample of 26 bright stars does not show a metallicity spread \citep{roederer16}. Instead they find large internal dispersion of [Mg/Fe] of 0.28 dex, which is akin to massive metal poor GCs such as $\omega$ Cen. The most recent study, using high-resolution spectra of 117 giant-stars of NGC 5824 \citep{mucci18}, confirms the lack of spread in [Fe/H] and the unusually large spread in [Mg/Fe]. It is also worth noting that deep photometric data from MegaCam and Dark Energy Camera \citep{walker17, kuzma18} find no evidence of tidal tails around NGC 5824, although the latter study does find that it has a very extended envelope with the radius of 230 pc. Such a feature may also be generated from the tidal stripping of a progenitor dwarf galaxy.

In this study we identify substructures in the outer stellar halo by crossing matching spectroscopic surveys with Gaia DR2. More specifically, we focus on the identified CS and its possible association with NGC 5824. The details of our catalog are described in Sec.~\ref{sec:data}. The methodology of the group identification and the map of the MW GCs is presented in Sec.~\ref{sec:med}. We show the identified streams and their associated GCs in Sec.~\ref{sec:res}. The conclusions and discussions are presented in Sec.~\ref{sec:con}.

\section{Data}
\label{sec:data}


For our analysis we combine two spectroscopic catalogs to create our parent sample. Firstly we use a catalog of K giants identified using a support vector machine classifier from LAMOST DR5 \citep{liu14}. We augment this using a catalog of BHBs from SDSS/SEGUE \citep{xue08, xue11}. We cross match these stars with Gaia DR2 to obtain proper motions, resulting in a final parent sample of around 100,000 stars.


Since we focus on the distant halo in this work, Gaia parallaxes are not suitable. Instead we obtain distance estimates from multi-band photometry \citep{xue08, xue14} and, in the case of the K-giant stars, spectroscopic metallicity from LAMOST. The mean uncertainties in distances are about 13$\%$ and 5$\%$ for K-giant stars and BHB stars, respectively. Using the spectra of K giants, the LAMOST pipeline gives line-of-sight velocities with uncertainties of around 5 -- 20 km s$^{-1}$ \citep{wu11, wu14}. There is a known systematic offset in the LAMOST radial velocity zero point. This has been quantified for LAMOST DR5 by \citet{tian18}, who compared to Gaia DR2 RVS velocities to obtain an offset of 5.4 km s$^{-1}$. We add this to our LAMOST radial velocities. Note that unlike previous versions of the LAMOST catalog, the DR5 pipeline has an improved algorithm for estimating uncertainties. This means that prior issues with overestimated uncertainties \citep[e.g.][]{schonrich17} have been alleviated and the uncertainties are now believed to be robust. For BHBs, line-of-sight velocities are derived from the SEGUE Stellar Parameter Pipeline (SSPP), which delivers uncertainties of 5 km s$^{-1}$ to 15 km s$^{-1}$ \citep{xue08}. Both pipelines provide measurements of [Fe/H] to an accuracy of around 0.1 to 0.2 dex\footnote{Note that the BHB metallicities in \citet{xue08} were obtained using a bespoke version of the SSPP.}.When we study the metallicity distributions of our identified halo groups, unless otherwise stated, we only include those members with S/N $>$ 20.


Since we are interested in streams and overdensities that are not yet phase-mixed, we focus our study on the outer halo, retaining only stars with helio-centric distances greater than 15 kpc. In order to minimize the contribution from the Sagittarius stream we further reject stars with [Fe/H] $>$ -1.5 dex. This cut should not remove a significant number of halo groups because we believe most accreted halo structures will come from dwarf galaxies, which typically have metallicities below -1.5 dex.
The final sample has 5,256 stars with 6-D phase-space information and [Fe/H]. For our selected K giants, $G$-band magnitudes from Gaia DR2 \citep{gaiadr2} are in the range 14 -- 18 mag, which have corresponding proper motion uncertainties in the range 0.04 -- 0.28 mas yr$^{-1}$. Therefore, taking into account our distance uncertainties of around 15\%, this means that typical transverse velocity uncertainties are around 10 km s$^{-1}$. The proper motion uncertainties of the final selected BHBs are slightly larger, at around 0.13 -- 0.6 mas yr$^{-1}$, owing to their fainter magnitudes ($G$ $\sim$ 17 -- 19 mag). This yields typical transverse velocity uncertainties of 20 km s$^{-1}$. After applying the above cuts, the radial velocity uncertainties are around 5 to 10 km/s for both populations. The uncertainty in each of the six phase-space dimensions is taken into account when we apply our group identification procedure (see details in Sec.~\ref{subsec:gi}).

\begin{figure*}[tb]
\centering
\includegraphics[width=\linewidth]{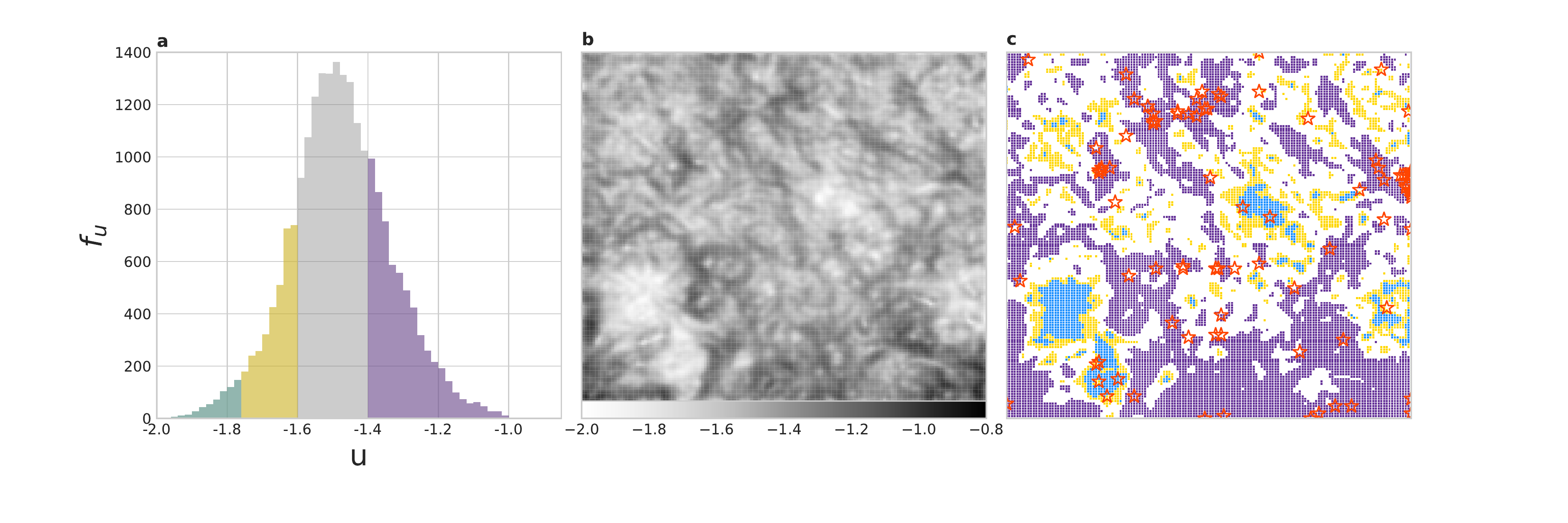}
\caption{Results from the application of \textsc{StarGO} to the metal poor outer halo catalog in the normalized ($E$, $L$, $\theta$, $\phi$) space. (a) The histogram shows the distribution of $u$ values between adjacent neurons, where the light blue and gold regions denote neurons with $u\leq u_{5\%}$ and  $u\leq u_{16\%}$ respectively. The purple region denotes neurons with $u>u_{60\%}$. (b) The 2D neuron map resulting from the SOM, where $u$ is represented by the gray color scale. (c) The same map as (b), but where neurons with $u\leq u_{5\%}$, and  $u\leq u_{16\%}$ are marked by light blue and gold pixels (these form the candidate groups) and those with $u > u_{60\%}$ are marked by purple pixels (these form the boundaries between candidate groups). The MW GCs that can be tightly associated with the neuron map are denoted by red stars.}
\label{fig:som}
\end{figure*}

\section{Method}
\label{sec:med}


Before going into the details of our approach, we briefly introduce our new cluster identification method \textsc{StarGO} \citep{yuan18}, which combines a self-organizing map (SOM) and a hierarchical group identification algorithm. The SOM trains the neural network to learn the data structure from an n-D input space and project it onto a 2D map. The clustering signatures within the n-D data set can be extracted from the visualization of the 2D neuron map. This is realized by introducing a weight vector for each neuron of the network, which has the same dimension as the input vector. The unsupervised learning is performed by iteratively updating the weight vectors so that they become closer to the input vectors from the data set. This process is complete when all of the weight vectors reach convergence. The trained neural network is visualized by the difference in weight vectors among adjacent neurons, which is denoted by $u$. Neighboring neurons with $u < u_{\mathrm{thr}}$ are identified as groups, where $u_{\mathrm{thr}}$ is determined based on the distribution of $u$, as well as the topological structure exhibited by the neuron map (see Sec.~\ref{subsec:gi} for a demonstration of how this works in practice). After the group identification procedure, we associate each star to its best matching unit (BMU) on the 2D map, which is defined to be the neuron with the weight vector closest to the input vector of that star. Stars associated with the identified neuron groups form the corresponding star groups.
The performance of \textsc{StarGO} in relation to the commonly used Friends-of-Friends algorithm has been quantified in \citet{yuan18}. 

\subsection{Input Space}
\label{subsec:input}

In this work we apply \textsc{StarGO} to the metal poor halo catalog using an input space constructed from angular momentum and orbital energy. More specifically, we use total angular momentum $L$, orbital energy $E$, and two parameters, $\theta$ and $\phi$, to characterize the direction of the angular momentum vector.
Our goal is to find star groups which are clustered in the ($E$, $L$, $\theta$, $\phi$) space. As we know, in an axisymmetric potential, $E$, $L$, and $L_z$ are approximately conserved and the latter two give $\theta = \mathrm{arccos}(L_z/L)$.
Although $\phi = \mathrm{arctan}(L_x/ L_y)$ is not strictly conserved, it has been shown that $L_x$ and $L_y$ evolve coherently \citep[e.g.][]{helmi00, knebe05, klement10, gomez10a, maffione15} and so we take this angle as the fourth dimension of our input space.
For the halo stars of extra-galactic origin, the exchange in angular momentum and orbital energy can be severe when their progenitors fall into the MW. However, stars from the same progenitor will have a similar disruption history, leading to similar changes in angular momentum and orbital energy.
Although the conservation of angular momentum and orbital energy will be violated, stars from a given progenitor can remain clustered in this space, as has been shown in mock simulations of substructures stripped from merging satellites \citep{yuan18}.
The final step is to make these quantities dimensionless, i.e. $(E-E_{\rm{min}})/E_{\mathrm{norm}}$, $(L-L_{\rm{min}})/L_{\mathrm{norm}}$, $\theta/\pi$, and $\phi/2\pi$, where $E_{\mathrm{norm}}$ and $L_{\mathrm{norm}}$ denote the ranges of the entire sample (e.g. $E_{\rm{norm}} = E_{\rm{max}} - E_{\rm{min}})$.

In a following paper, Xue et al. (in prep) will present a companion analysis using a Friends-of-Friends approach, based on a similar input space.

\subsection{Group Identification}
\label{subsec:gi}

We feed the input data into a 150$\times$150 neural network, using 400 iterations of the learning process. 
Since we use two angles to represent the direction of the angular momentum, we use a great-circle distance to calculate the difference between neurons.
Therefore in our final trained map the difference between neighbouring neurons $i$ and $j$ (hereafter $u_{i,j}$) is defined by the following equations,

\begin{equation}
I_{i,j} = \mathrm{sin}(\delta\theta_{i,j}/2)^2+\mathrm{cos}\theta_i\mathrm{cos}\theta_j \mathrm{sin}(\delta\phi_{i,j}/2)^2, 
\end{equation}

\begin{equation}
\begin{split}
u_{i,j} & = \log_{10}[(\delta E_{i,j}/E_{\mathrm{norm}})^2 + (\delta L_{i,j}/L_{\mathrm{norm}})^2 \\
& + (2/\pi\cdot\mathrm{arcsin}\sqrt{I_{i,j}})^2].
\end{split}
\label{eq:uij}
\end{equation}

The trained neural network is visualized by the gray-scale map of $u$ (see Fig.~\ref{fig:som}b), where $u$ represents the difference in weight vectors between adjacent neurons.
The clouds in the gray-scale correspond to distinctive structures within the underlying data set, with darker shades denoting higher values of $u$, i.e. steeper gradients in the weight vectors.

We first mark neurons with $u<u_{5\%}$ as light blue pixels and neurons with  $u<u_{16\%}$ as gold pixels in Fig.~\ref{fig:som}c, where $u_{5\%}$ and $u_{16\%}$ denote the 5th and 16th percentile of the distribution of $u$ respectively (see Fig.~\ref{fig:som}a).

The light blue patches of neurons are very similar (among the top 5\% of all neurons) in the weight vector space. They are connected by the gold neurons, which are the top 16\% of all the neurons. These highlighted neurons reveal the clustering structures from the input space. The inner dense cores of clusters are associated with the light blue neurons, and the less dense outskirts are associated with the gold neurons. In the trained map we can easily see three regions that are populated by significant light blue patches connected by gold neurons: one in the middle, one in the lower left corner, and one in the lower right corner. We consider these groups as candidate groups, and try to identify their group members systematically.

We then mark the neurons with the lowest clustering significance as purple pixels, where we use a threshold $u>u_{\mathrm{thr}}$. This threshold parameter $u_{\mathrm{thr}}$ is adjusted in order to maximize the size of each group, as described below. For Fig.~\ref{fig:som}c we have chosen $u_{\mathrm{thr}} = u_{60\%}$, i.e. the purple pixels correspond to neurons with clustering significance in the bottom 40\%. 
These pixels form the boundaries separating candidate neuron groups into isolated islands, such as the islands corresponding to the salmon and cyan candidate groups in Fig.~\ref{fig:gi}Ia, IIa.
Neurons within the islands have similar weight vectors and these weights are significantly different from the neurons in the purple boundary regions. These islands represent the strongest clustering signals in the underlying 4-D data set. 

In order to systematically identify clustered star groups, we start with $u_{\mathrm{thr0}} = u_{100\%}$. At this stage our neuron map consists of one large group. As we gradually decrease the value of $u_{\mathrm{thr}}$, boundary (purple) pixels begin to populate the map and separate islands begin to emerge (analogous to the rising tide cutting off Mont-Saint-Michel from the mainland). The value of $u$ for which the first significant island emerges is denoted as $u_{\mathrm{thr1}}$, which in this case corresponds to $u_{66\%}$. This produces the salmon island in Fig.~\ref{fig:gi}Ia.
We then continue decreasing $u_{\mathrm{thr}}$ until the next large island forms, which is at $u_{\mathrm{thr2}} = u_{60\%}$ (corresponding to the cyan island in Fig.~\ref{fig:gi}IIa). As we keep decreasing $u_{\mathrm{thr}}$  
the island in the lower right corner appears when $u_{\mathrm{thr3}} = u_{30\%}$ (the magenta island in Fig.~\ref{fig:gi}IIIa). At this threshold, the region populated by light blue patches in the middle of the map is still connected with the gold patches in the upper right corner (Fig.~\ref{fig:gi}IIIa). We could continue the above prescription to isolate this candidate group and other, smaller groups. We defer the analysis of these new groups to a subsequent paper.


Observational uncertainties are taken into account by utilizing the 2-D neuron map. For each star we generate 1000 Monte Carlo realizations, given each star's uncertainty in distance, radial velocity, and proper motion. We assume all uncertainties follow Gaussian distributions.
After the creation of these 1000 realizations, for each one we identify its BMU in the trained map. Note that the BMU for each realization is not necessarily the same as the BMU of the fiducial star.

In principle, any star can be mapped to the trained neural network. However, the difference in the input vector and the weight vector of its BMU can be very large if the input vector is very different from the training sample. Mapping such stars has little meaning since they are not closely associated to any neurons. Thus we need to define the limit of the trained neural network, within which the mapping can be considered as appropriate. We first calculate the 4-D distance between stars in the training sample and their BMUs, which is referred to as u$_{\mathrm{vw}}$. Then we can get the largest distance denoted by u$_{\mathrm{vw}, max}$ out of the entire sample. This maximum offset denotes the limits of the neuron map, i.e. stars that lie beyond this limit cannot be closely attached to any neuron of the map. We remove such data points from our set of realizations if the 4-D distance to its BMU is larger than u$_{\mathrm{vw}, max}$.

We also wish to test whether the star's association to its respective group is robust. After rejecting realizations according to the preceding paragraph, we then analyze the BMU for each remaining realization. For illustrative purposes, we plot the 2-D density map of all the remaining realizations for the magenta group in Fig.~\ref{fig:err}a. It can be seen that some BMUs lie outside the boundary of the identified group and we consider these as invalid realizations.

After we have carried out these two steps and obtained the total number of valid realizations for each of the candidate members, we reject the worst 16\%, where ``worst" corresponds to the candidate members with the fewest valid realizations.
This sets the confidence level of member identification for each group individually. 
For OS the candidate members all have at least 200 valid realizations, which means more than 200 out of 1000 realizations for each member in OS are associated with the corresponding neuron group. For CS and Sgr the candidate members all have at least 400 valid realizations out of 1000.

\subsection{Mapping the globular clusters of the Milky Way}
\label{subsec:gc}

\begin{figure}[ht]
\centering
\includegraphics[width=\linewidth]{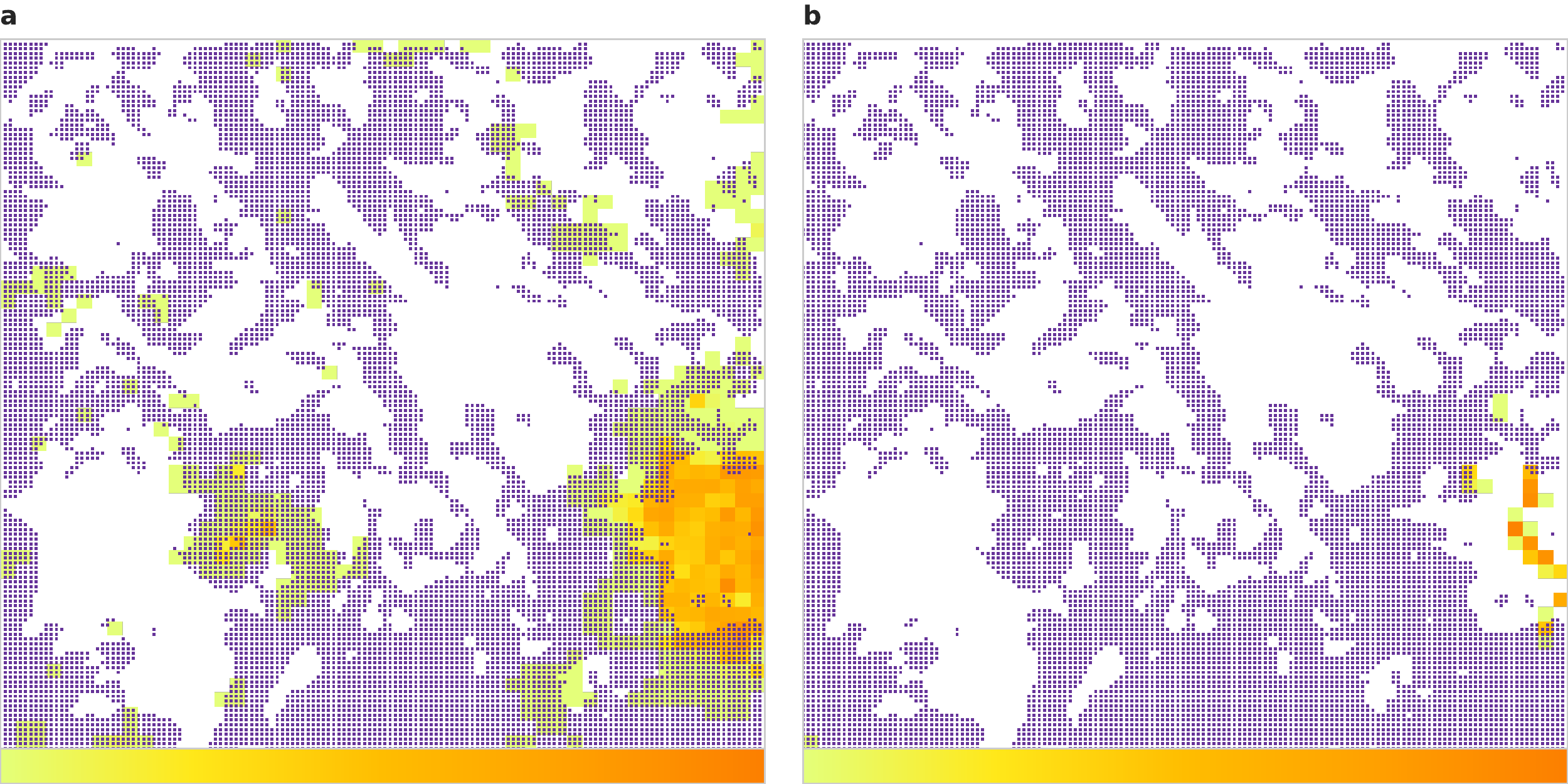}
\caption{
(a) The background purple distribution shows the same neuron map as Fig.~\ref{fig:gi}IIIa, overlaid with an orange heatmap corresponding to the density of BMUs associated to the mock stars resampled from candidate CS members. Most of the BMUs lie within the identified CS group (in the lower-right corner), although a small fraction lie outside this group. Candidate members that have a significant fraction of resampled mock stars lying outside the group are not considered as valid members.
(b) As (a), but here we plot the mock realizations of NGC 5824, showing that almost all realizations lie within the identified CS group.}
\label{fig:err}
\end{figure}
\begin{figure*}[tb]
\centering
\includegraphics[width=\linewidth]{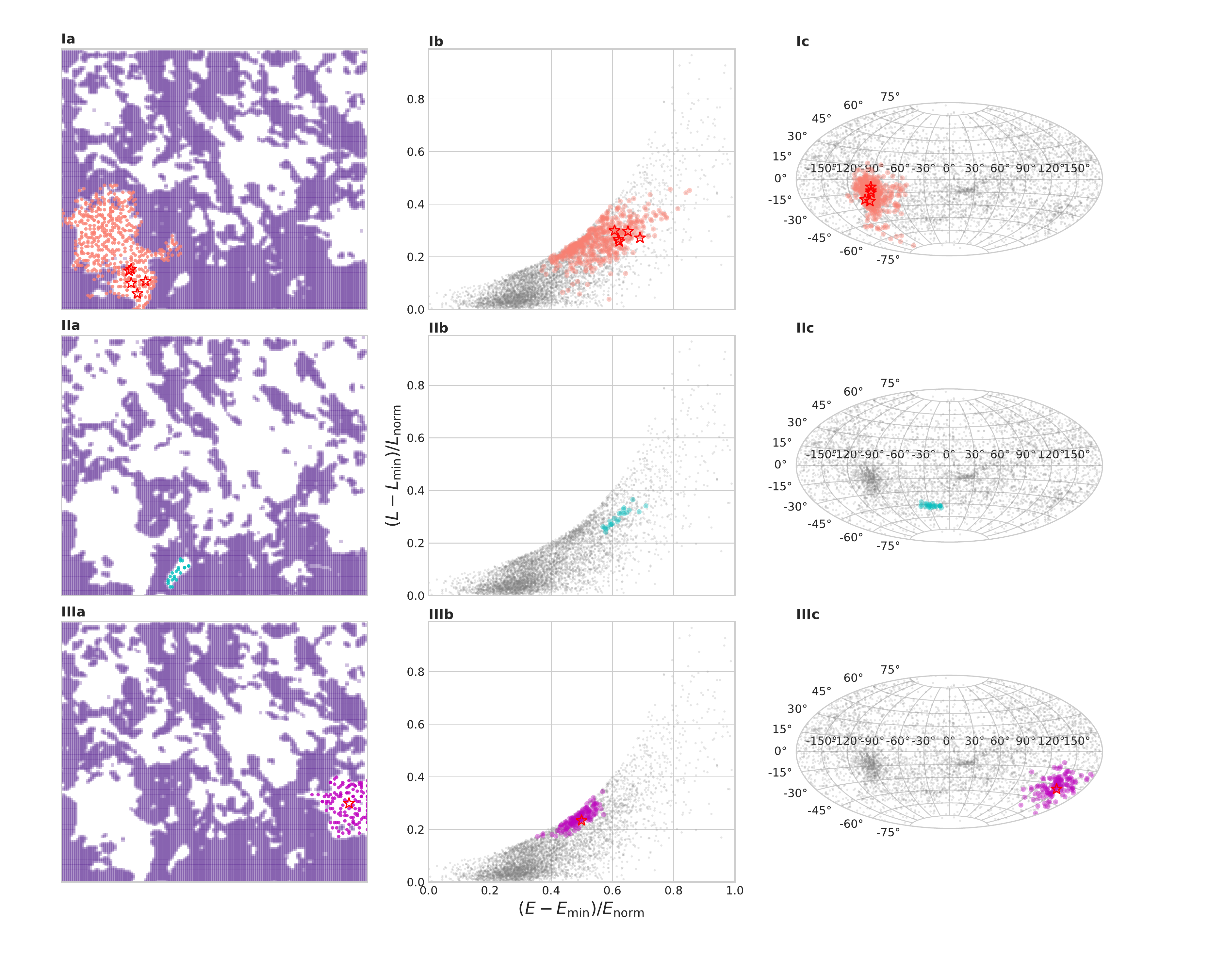}
\caption{Group identification results from \textsc{StarGO}. (Ia) 
This panel shows the same SOM as in Fig.~\ref{fig:som}c, but with a threshold of $u_{\mathrm{thr1}}=u_{66\%}$ chosen to identify the first group (salmon points). Boundary neurons are colored purple. The properties of the group are shown in panels (Ib) and (Ic), where the entire metal-poor halo catalog are shown in grey and the identified group members are shown in salmon. Panel (Ib) shows the distribution of normalized energy and angular momentum, while panel (Ic) shows the orientation of the angular momentum vectors in a Galacto-centric projection. The second and third rows correspond to the other identified groups with $u_{\mathrm{thr2}}=u_{60\%}$ (cyan) and $u_{\mathrm{thr3}}=u_{30\%}$ (magenta), respectively. The red stars denote MW GCs that are associated to these groups.}

\label{fig:gi}
\end{figure*}

In order to study the relationship between the MW GCs and the identified star groups, we attach each GC to their BMU on the trained neuron map
This process requires the kinematics of each GC in order to calculate its angular momentum and orbital energy. These kinematics are taken from the Harris catalog \citep{harris96, harris10} and \citet{gc18}. We normalize the input vector of each GC by $E_{\mathrm{norm}}$ and $L_{\mathrm{norm}}$ from the training sample (see Section \ref{subsec:input}). We also exclude all GCs for which the 4-D distance to the BMU is larger than the largest offset between a star in our training sample and its BMU. The GCs that pass these cuts are mapped to their BMUs and denoted by red stars in Fig.~\ref{fig:som}c, where is can be seen that many lie clustered in this space and are associated to over-densities in the input stellar halo catalog. We plot the GCs attached to the magenta and salmon islands in Fig.~\ref{fig:gi}I, III as red stars. Following the same recipe as the previous section, for each GC that is associated to one of the groups we generate 1000 realizations to test for robustness. We find that each cluster has at least 900 valid realizations.
We illustrate this process in Fig. \ref{fig:err}b for the GC lying in the magenta island, showing the 2-D density map of all realizations that can be closely attached to the neuron network.

\section{Results}

\label{sec:res}

\begin{figure*}[tb]
\centering
\includegraphics[width=\linewidth]{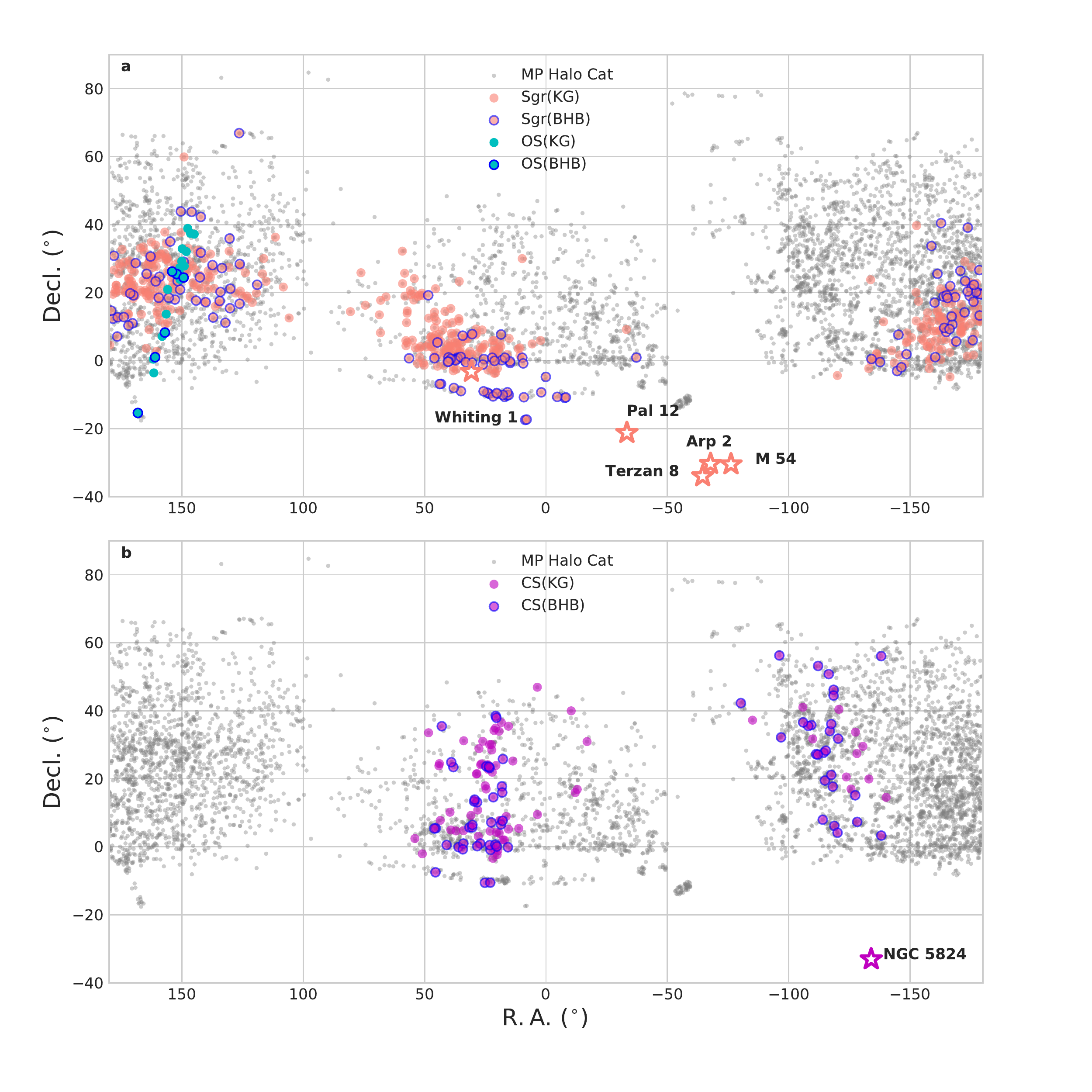}
\caption{The identified streams in equatorial coordinates. The metal-poor outer-halo catalog is shown in gray. (a) The valid members of OS and Sgr are plotted with solid cyan and salmon filled circles, respectively. BHB members have a blue border and K-giant members have no border. The GCs associated with Sgr are denoted by salmon stars and labelled. (b) As (a), but for CS. The associated GC (NGC 5824) is denoted by the magenta star.}
\label{fig:streams}
\end{figure*}

\begin{figure}[ht]
\centering
\includegraphics[width=\linewidth]{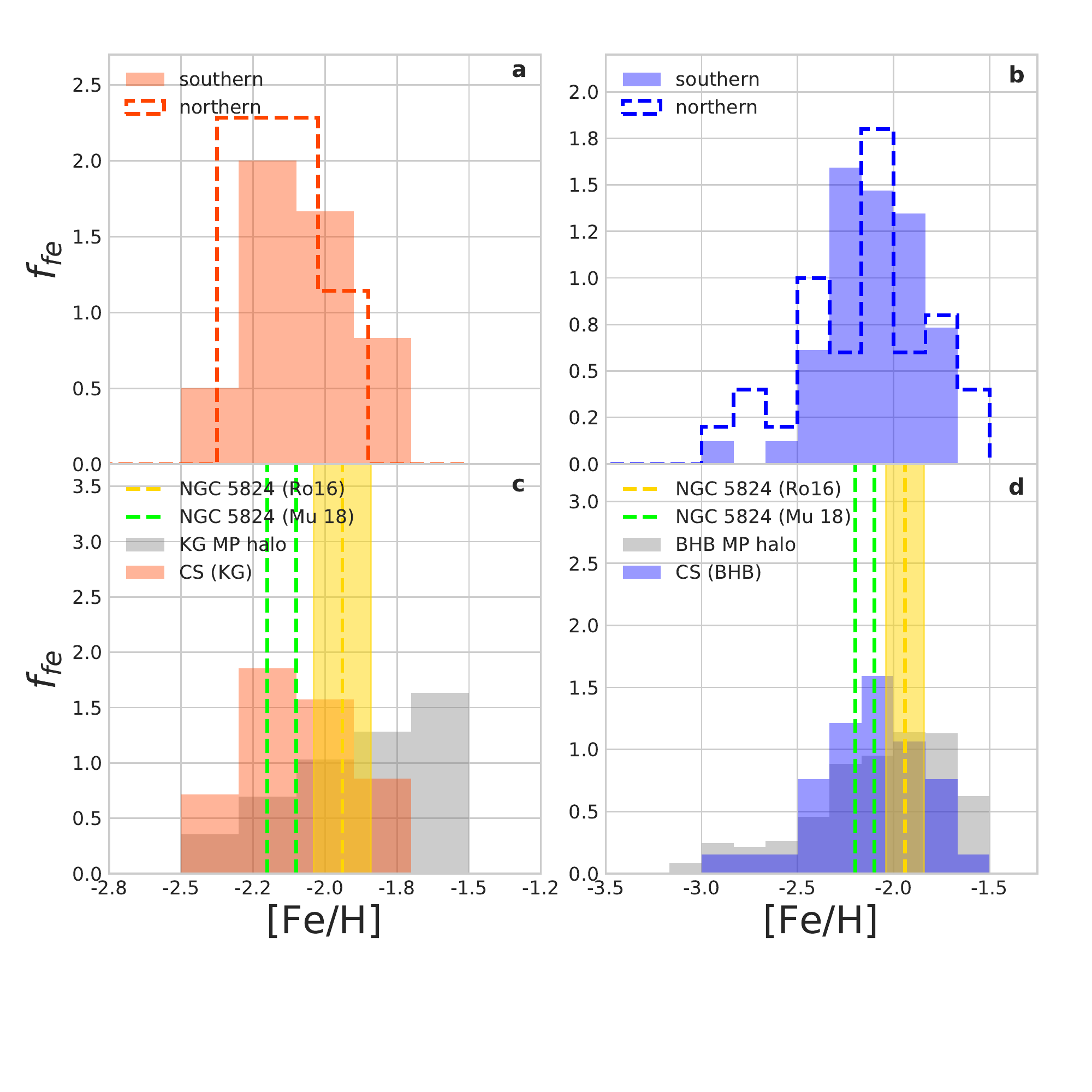}
\caption{The filled and dashed histograms represent the normalized metallicity distributions of the southern and northern counterparts of the CS, respectively, for K-giants (left panels), and BHBs (right panels). The lower panels show the combined distributions of both northern and southern counterparts. The gray histograms represent the distribution of the entire halo sample, clearly showing how the CS members are homogeneous and inconsistent with the background population (especially for the K-giants). In these panels the gold dashed lines and shaded bands shows the mean [Fe/H] for NGC 5824 and its uncertainty, respectively, according to \citet{roederer16}. The lime dashed lines represent the mean metallicities derived by \citet{mucci18} for red giant branch stars (-2.11 dex) and asymptotic giant branch stars (-2.20 dex).}
\label{fig:met}
\end{figure}


The first iteration of \textsc{StarGO} generates the 2D neuron map shown in Fig.~\ref{fig:som}, which was introduced in Section \ref{subsec:gi}. As discussed in Section \ref{subsec:gi}, our analysis of this map results in three identified groups, which we will now discuss. These are illustrated in Fig.~\ref{fig:gi}, where each row corresponds to one of the identified groups.

\subsection{The Orphan and Sagittarius streams}
\label{subsec:os_sgr}

In this subsection we demonstrate the efficacy of our method by presenting two groups that correspond to previously-identified halo streams.

The first group identified in Section \ref{subsec:gi} is shown in the first row of Fig.~\ref{fig:gi}. It is the largest identified indivisible group in our halo catalog, consisting of 390 K giants and 132 BHBs. From its sky position (Fig.~\ref{fig:streams}) we can easily see that this group corresponds to the Sagittarius stream \citep[Sgr; e.g.][]{maj03,belokurov06}. As discussed in Section \ref{subsec:gc}, we have mapped a sample of GCs onto the same neural network. We find that five of these GCs are closely associated to our Sagittarius group (the red stars in Fig.~\ref{fig:gi}). These are Whiting 1, NGC 6715 (M 54), Arp 2, Terzan 8, and Pal 12, all of which have been previously associated to the Sagittarius stream \citep{law2010a, law2010b, sohn18}.

Next we discuss the group in the lower-left corner of our neuron map, shown in the second row of Fig.~\ref{fig:gi}. Although there are only 22 stars belonging to this group, it is compact in both energy and angular momentum spaces. The sky position of these stars are shown in Fig.~\ref{fig:streams}, from which it is clear that this group corresponds to the Orphan Stream \citep[OS; e.g.][]{orphan06}. The distances to these stars range from 18 kpc to 47 kpc, with a clear gradient along the stream, i.e. similar to the results from SEGUE \citep[19 -- 47 kpc;][]{orphan10}. The OS is also detected in the southern sky to Decl. = -38 deg \citep{grillmair15}. Recent study by \citet{koposov19} shows an all-sky view of the OS, which has $\sim$ 210 deg length. However we can only identify its stream members with Decl. $>$ -20 deg as our halo catalog consists of data from northern hemisphere surveys (SDSS/SEGUE and LAMOST).

\begin{figure*}[hb]
\centering
\includegraphics[width=\linewidth]{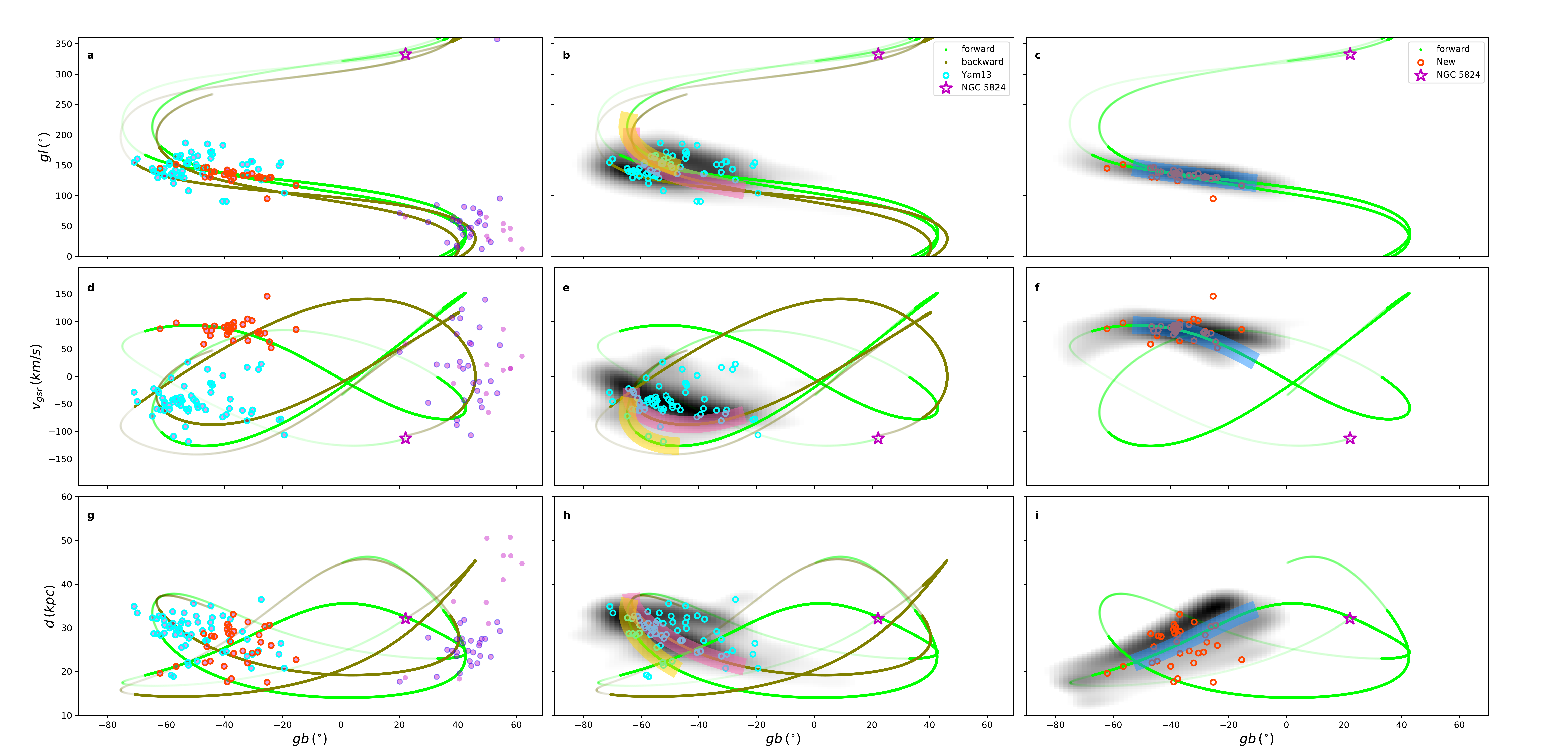}
\caption{Various properties of the identified CS members plotted as a function of Galactic latitude: Galactic longitude (top row), radial velocity in the Galactic rest frame (middle), and helio-centric distance (bottom). Magenta points denote the CS members, and the magenta star denotes NGC 5824. The southern detection is bifurcated into two components with positive (red circles) and negative (cyan circles) radial velocities. This latter component is consistent with the detection in \citetalias{yam13}. The green/olive lines denote the orbit of NGC 5824 integrated for 1.8 Gyr ($\sim$ 3 orbital periods) forwards/backwards, where the thin lines denote parts of the orbit for which ${\rm Decl.} < -10$ deg, i.e. these cannot correspond to our debris, as our stars are all located at ${\rm Decl.} > -10$ deg. In the middle column, the grey-scale map denotes the orbits of the cyan stars (\citetalias{yam13}), integrated both forward and backwards for 0.1 Gyr in order to show their gradients. Note that these stars match well two portions of the orbits of NGC 5824. One portion is around 1.5 Gyr behind the cluster (transparent pink band) and the other is about 1.2 Gyr ahead (transparent gold band). In the right column, the grey-scale map denotes the orbits of the red stars integrated forward for 0.1 Gyr. This new component matches with the portion of the orbit of NGC 5824 around 0.3 Gyr ahead of the cluster (transparent blue band).}
\label{fig:orbit}
\end{figure*}

\subsection{The Cetus Stream and NGC 5824}
\label{subsec:cps}

The final of our three groups is shown in the third row of Fig.~\ref{fig:gi}. This contains 151 members from our halo catalog and, as with the other groups, it is highly clustered in the input space (see Fig.~\ref{fig:gi}IIb,IIc). The sky position of these members shows two separated regions, one in the southern Galactic hemisphere (at 0 $\ga$ R.A. $\ga$ 50 deg, corresponding to $l \sim 150$ deg, $b \sim -40$ deg) and one in the north (at -150 $\ga$ R.A. $\ga$ -100, corresponding to $l \sim 50$ deg, $b \sim 50$ deg). The former region is the most dense, with 106 stars. The distance (20 -- 35 kpc) and sky position tells us that this corresponds to the Cetus stream \citep[CS; e.g.][]{newberg09, yam13}. Although it overlaps with part of the Sgr stream in sky position (see Fig.~\ref{fig:streams}) and has a similar distance, the directions of the angular momentum of Sgr and CS are clearly different (Fig.~\ref{fig:gi}Ic,IIIc). 

While the detection in the southern Galactic hemisphere is not new, the detection of the northern counterpart has not previously been found. This new feature consists of 45 stars and is more diffuse on the sky. Because both counterparts come from the same single identified group in the neuron map, sharing the same total energy and angular momentum, we believe these two overdensities are from the same stream. We hereafter refer to these as the northern and southern counterparts of the CS.

\subsubsection{Statistical significance of the northern detection}

Before we discuss the association of these two detections, we first test the significance of the new northern counterpart. This is done using a smooth halo drawn from the mock Gaia DR2 catalog of \citet{mock18}, which is based on the popular \texttt{Galaxia/Besancon} models \citep{galaxia,besancon}. We create a sample of mock halo stars similar to our observed sample of metal-poor stars, taking halo stars (i.e. age $\geq$ 11 Gyr) with $\rm Decl. \ga -10$ deg, d $\ga$ 15 kpc, [Fe/H] $<$ -1.5 dex, and $G$ band magnitude $<$ 19. We then feed these mock stars through our neural network and select those which are tightly associated to the CS detection (i.e. the magenta region in Fig. \ref{fig:gi}IIIa). We find that the smooth halo produces 3,301/155,343 = 2$\%$ stars associated to the CS detection, which is far fewer than the observed fraction of 151/1,679 = 9$\%$. If we concern ourselves with only the new northern detection (restricting ourselves to $b\geq$20 deg), we find that the mock sample has 1,795/107,486 = 1.6\% of stars associated to this detection, whereas the observed sample contains 45/1,105 = 4\%. In order to put this on a more quantitative footing, we can calculate the corresponding binomial probability, i.e. the probability that a smooth halo can, by chance, provide 45 stars which lie within our CS detection region. This takes into account the Poisson fluctuation in the background population, under the assumption that the sample of 1,105 halo stars is drawn from a smooth population. We find that the binomial probability of obtaining 45 stars in our detection region, from a sample of 1,105 smooth stars, is less than 1e-7\%, i.e. the significance is more than 5-sigma.

\subsubsection{Metallicity}
 
In Fig.~\ref{fig:met} we plot the metallicities of our CS members. We divide our sample into K-giants($g$-band S/N $>$20), and BHBs, for both northern and southern detections. The metallicities of the K-giants confirm our claim that these two detections are part of the same stream. Both northern and southern detections have very similar metallicities (see Fig.\ref{fig:met}a), and crucially these metallicities are clearly distinct from the background halo population (see Fig.\ref{fig:met}c). If we apply a KS test to assess the probability that the metallicities of the northern and southern detections are consistent with the background halo population, we obtain p-values below 0.01, i.e. they are clearly inconsistent. Note that the K-giant metallicities have been calculated using the LAMOST metallicity pipeline, which only extends to around -2.5 dex \citep{luo15}, and so the lack of a low-metallicity tail compared to the BHBs is not surprising.

We measure the mean and dispersion of the metallicity for these stars using a Bayesian maximum likelihood technique \citep{emcee}, taking into account the observational uncertainties of around 0.2 dex. We find that the BHBs have a mean of $-2.15 \pm 0.04$ dex and a dispersion of $0.23 \pm 0.04$ dex, while the K-giants have a mean of $-2.07 \pm 0.02$ dex and a dispersion of $0.12 \pm 0.02$ dex. The means are consistent and, although the dispersions of the two populations differ slightly, it is clear that these dispersions are inconsistent with zero at a high level of significance. We therefore conclude that the stream comes from a disrupted dwarf galaxy, rather than a disrupted globular cluster for which we would expect a dispersion of $\sim$ 0 dex. We return to this issue in the discussion. 

\subsubsection{Association to NGC 5824}

As before, we check to see whether any MW GCs could be associated to this group and find that NGC 5824 (the red star in Fig.~\ref{fig:gi}III) is tightly associated. As shown in Fig. \ref{fig:err}b and discussed in Section \ref{subsec:gc}, our Monte Carlo analysis demonstrates that the association is robust. We defer a discussion of the orbital properties of the cluster to the following subsection.


We also compare the metallicity of CS members with high-resolution studies of NGC 5824. This is shown in Fig.~\ref{fig:met}cd. \citet{roederer16} found [Fe/H] = -1.94$\pm$0.12 dex from 26 red giants stars (see gold dashed line with transparent band). In a later study, \citet{mucci18} found [Fe/H] = -2.11$\pm$0.01 dex (right lime dashed line) from 87 red giant branch stars, and [Fe/H] = -2.20$\pm$0.01 dex (left lime dashed line) from 30 asymptotic giant branch stars. Overall, the typical metallicity of CS members agrees very well with that of NGC 5824 from these high resolution spectroscopic studies. This fact supports the scenario that they are from the same progenitor system. We return to this issue later in the discussion.

\subsubsection{Orbital properties}

We now investigate the orbit of the stream. Hitherto the orbit has not been well-determined, owing to the lack of precise proper motions \citepalias[e.g.][]{yam13}, but Gaia DR2 enables us to analyze the orbit in much greater detail. Fig. \ref{fig:orbit} shows the properties of our stars in terms of helio-centric distance and radial velocity as a function of Galactic latitude, which is chosen as the proxy for angle along the stream because the orbit has a large extension in Galactic latitude.

We start by analyzing the stars from the existing detection in the south ($-70 \la b \la -20$ deg). We can see from the Galactic rest frame radial velocity distribution ($v_{\rm gsr}$) that our detection appears to be split into two clumps, one with $-75 < v_{\rm gsr} < -25$ km/s (cyan circles in Fig.~\ref{fig:orbit}) and one with $50 < v_{\rm gsr} < 100$ km/s (red circles in Fig.~\ref{fig:orbit}). The former clump corresponds to the detection in \citetalias{yam13}, which has a negative radial velocity gradient with $b$. The clump at positive $v_{\rm gsr}$ is new, although it should be noted that \citetalias{yam13} found a number of stars with velocities that are consistent with this clump. Due to small number statistics they were unable to make the association, but with our powerful \textsc{StarGO} tool and a much larger catalog from LAMOST we are now able to do so. What we appear to be detecting is phase mixed material from the progenitor system.

To understand this better, we compare the properties of the stream to the orbit of NGC 5824, which has pericenter and apocenter at 19 kpc and 41 kpc respectively.It is possible that they are stripped from the same progenitor system. In this scenario, even though tidal streams do not necessarily follow the precise orbit of the progenitor, it should still enable us to gain a clearer picture of its disruption. Fig. \ref{fig:orbit} shows orbit of the cluster integrated forward and backwards for 1.8 Gyr, which is around three times its orbital period ($T=0.6$ Gyr). The thin lines denote parts of the orbit for which ${\rm Dec} < -10$ deg, i.e. these cannot correspond to our debris, as our stars are all located at ${\rm Dec} > -10$ deg.

We first focus on the existing detection from \citetalias{yam13}, given by the cyan circles in Fig. \ref{fig:orbit}. The gray-scale maps in the middle columns show the orbits of these cyan stars integrated for a short time in order to illustrate their gradients. From these, there are two portions of the NGC 5824 orbit that are consistent with this debris, highlighted by the transparent gold and pink bands. The former is around 1.2 Gyr ahead of the cluster ($\sim 2$ orbital periods), while the latter is around 1.5 Gyr behind ($\sim 2.5$ orbital periods). The new detection at positive radial velocities (red points) is shown in the right columns. Here there is only one portion of the orbit that is consistent with the debris, highlighted with the transparent blue band. This is around 0.3 Gyr ahead of the cluster ($\sim 0.5$ orbital periods). The new detection in the northern Galactic hemisphere (magenta points) is harder to discern due to the wide spread in radial velocities, but is consistent with a number of portions of the orbit. Note that even though the radial velocities cover a wide range, their angular momenta are all tightly concentrated (see Fig. \ref{fig:gi}).

If the above picture is correct, it is curious as to why there are gaps in the detecting material. Part of this can be explained by portions of the orbit that lie outside our field of view (either at negative declinations or behind the Galactic plane), but there are other portions which we could be expected to detect. In particular the trailing material from less than one orbital period behind the cluster should be visible at $-70 \la b \la -20$ deg. This portion of the stream corresponds to pericentre of the orbit and so perhaps the stream is sparsely populated here.

From this orbital analysis, together with the homogeneous metallicities shown in Fig. \ref{fig:met}, we conclude that these new detections (both at positive $v_{\rm gsr}$ and in the northern hemisphere) 
have orbits that are consistent with the Cetus stream from \citetalias{yam13} as well as NGC 5824. It is very likely that all of this material comes from one disrupted system. The system leaves material at a wide range of orbital phases (about 4.5 orbital periods), and as a consequence the stream is very diffuse. 

It is possible that this system could consist of multiple components that fell in together with very similar orbital properties. A plausible picture is that NGC 5824 may come from another dwarf galaxy, which fell into the MW along with the progenitor dwarf of the Cetus Stream. In this case, the different portions of the stream may be stripped from these two galaxies. Group infall is the most-common scenario for the accretion of satellite galaxies in zoom-in simulations of MW-like systems \citep{wetzel15}. There is also observational evidence supporting this picture. For example, the Leo-Crater group \citep{dejong10, belokurov14, torrealba16} contains several low mass dwarf galaxies (Leo V, Leo IV, and Crater 2) and Crater Globular Cluster. Recently, \citet{kall18} showed that four newly discovered ultra-faint dwarf galaxies may have fallen in with the Magellanic Clouds.

\section{Conclusions and Discussions}
\label{sec:con}


In this work we have applied a novel clustering method \textsc{StarGO} to a metal poor halo catalog with full 6D phase-space information, constructed by cross-matching Gaia DR2 with spectroscopy from LAMOST DR5 and SDSS/SEGUE. After the first iteration of the workflow we are able to identify three significant star groups, two of which can be easily confirmed as the Sagittarius and Orphan streams. The third star group corresponds to the Cetus Stream (CS). Our detection populates two separate regions on the sky, one in the southern Galactic hemisphere and one in the north. The southern counterpart (at 0 $\ga$ R.A. $\ga$ 50 deg) corresponds to the previous CS from \citet{newberg09} and \citetalias{yam13}. The northern counterpart (at -150 $\ga$ R.A. $\ga$ -100 deg) is a new discovery and has the same angular momentum and orbital energy as the southern counterpart.

In order to understand more about the progenitor system, we have investigated the orbit of the CS in detail, which is now possible thanks to the precise proper motion information from Gaia. Despite often being referred to as the Cetus Polar Stream, we have found that the angular momentum vector of our CS members is not actually aligned with the Galactic plane, as expected for a polar stream, and is offset by around 45 deg (see Fig.~\ref{fig:gi}IIIc). We therefore suggest future studies drop the term ``polar" from its name.
We also found the southern counterpart of the CS bifurcates in radial velocity. The clump with negative radial velocity is close to the \citetalias{yam13} detection, which has a negative distance gradient with b. The new clump has a positive distance gradient with b and, when compared to the orbit of the existing material, aligns well albeit with a phase offset of around 0.9 Gyr (corresponding to $\sim$1.5 times the orbital period).

We believe that the new northern counterpart is a mixture of debris. In general the stream is very diffuse, as had been seen in previous studies. The fact means that it is hard to identify in configuration and/or velocity space. The high degree of phase mixing suggests the progenitor was massive and/or was stripped a long time ago \citep{johnston01,amorisco15}. A suite of N-body simulations are needed in order to test these possibilities and draw quantitative conclusions about the progenitor and its disruption history. 

We have checked the metallicity of the southern and northern parts and find that they are in very good agreement, noticeably offset from the background halo population. The intrinsic metallicity dispersion (i.e. after accounting for observational errors) of all the CS members from the BHB catalog is 0.23 $\pm$ 0.04 dex, which is significantly larger than the spread expected for a GC and more-consistent with a dwarf galaxy. According to the stellar mass - stellar metallicity relation for Local Group dwarf galaxies from \citet{kirby13}, the metallicity of $\sim$-2 dex suggests a stellar mass in the range 10$^4$M$_{\odot}$ -- 10$^6$M$_{\odot}$, i.e. similar to low-mass classical MW dwarf galaxies such as Ursa Major I or Ursa Minor. The identified CS has a remarkable high ratio of BHB to K-giant stars (81/70), as has been noted previously \citet{newberg09, koposov12, yam13}. The high fraction of BHBs suggest the progenitor system had most of its stars formed in an early period of star formation.


By exploiting the visualization power of \textsc{StarGO}, we are able to map MW GCs to the trained neural network. We find a strong dynamical association between the Sgr stream and five of its surrounding GCs, which has been reported in previous studies \citep{law2010b, sohn18} using both $HST$ and $Gaia$ DR2 data. For the CS, its possible relationship between NGC 5824 was initially postulated using radial velocities in its discovery paper \citep{newberg09}. Thanks to Gaia we can now obtain 6-D kinematics of both the CS and NGC 5824. From its angular momentum and orbital energy we find that NGC 5824 is located within the same group as the CS members, confirming the association between the cluster and the stream. By integrating the orbit of NGC 5824 we find that the stream is consistent with material stripped.

The metallicities of our CS detection, both for K-giants and BHBs, is in good agreement with that of NGC 5824 from spectroscopic studies of the cluster \citep{costa14,roederer16,mucci18}, providing independent evidence of their common origin. Another possible scenario is that NGC 5824 comes from a different dwarf galaxy progenitor that shares a similar orbit and mass with the progenitor of the Cetus Stream. This happens when a group of satellites fall into a large host galaxy such as the MW. Such a picture is supported by the group infall scenario from both observations and simulations.

NGC 5824 is also reported to have a small spread in [Fe/H] but large spread in [Mg/Fe] \citep{roederer16, mucci18}, which is typically found only in massive and/or metal-poor clusters such as $\omega$ Cen. The mass-luminosity relation for GCs suggests a stellar mass of over $10^6~{\rm M}_\odot$. This is comparable to, or even greater than, the stellar mass of the progenitor dwarf estimated above ($10^4$ to $10^6~{\rm M}_\odot$). High resolution hydrodynamical simulations have shown such systems can be robustly produced, with star clusters forming at the centers of some halos at high redshift \citep{kimm16,ricotti16}. Given the mass ratio between NGC 5824 and its host galaxy, if the star cluster is initially born off-center, it will inspiral into the center in a short timescale ($\sim$ 1 Gyr) due to dynamical friction \citep{binney08}. According to the quantitative comparisons between the MW GC population and cosmological zoom-in simulations from the E-MOSAICS project, the MW hosts 6$\pm$1 nuclear clusters from disrupted dwarf galaxies \citep{kruijsse19}. NGC 5824, like M54 and wCen, is likely to be the nuclei of an accreted dwarf galaxy.

Clearly this is a complex system. The fact that it appears to be almost fully phase-mixed means that orbital analyses (such as those we have presented) are always going to have limited power. Detailed N-body simulations are required to fully understand the history of the progenitor's accretion and its relation to NGC 5824. We have begun to undertake this task and will report our findings in due course.

\acknowledgments

This work has made use of data from the European Space Agency (ESA) mission
{\it Gaia} (\url{https://www.cosmos.esa.int/gaia}), processed by the {\it Gaia}
Data Processing and Analysis Consortium (DPAC, \url{https://www.cosmos.esa.int/web/gaia/dpac/consortium}). Funding for the DPAC
has been provided by national institutions, in particular the institutions
participating in the {\it Gaia} Multilateral Agreement.

The authors wish to thank Hao Tian and Jundan Nie for sharing the Match Filter code when checking the tidal tails of NGC 5824, and Jo\~ao A. S. Amarante for providing an introduction to emcee. Z.Y. and M.C.S are partly supported by the National Key Basic Research and Development Program of China (No. 2018YFA0404501) and NSFC grant 11673083. Z.Y. and J.L. acknowledge the support from Special Funding for Advanced Users through LAMOST FELLOWSHIP. J.L. also acknowledges the NSFC grant 11703019. X.-X Xue thanks the support of  ”Recruitment Program of Global Youth Experts” of China and NSFC under grants 11390371,11873052,11890694.  This project is developed in part at the 2018 Gaia-LAMOST Sprint workshop, supported by the National Natural Science Foundation of China (NSFC) under grants 11333003 and 11390372. Z.Y. and Y.W. are partly supported by China Postdoctoral Science Foundation. Y.W. also acknowledges the Young Researcher Grant of National Astronomical Observatories, Chinese Academy of Sciences; and the National Natural Science Foundation of China under grant 11803048. J.C. is supported by the Astronomical Big Data Joint Research Center, co-founded by the National Astronomical Observatories, Chinese Academy of Sciences and the Alibaba Cloud.

\bibliography{ms}
\bibliographystyle{apj}

\end{document}